# Deep learning-based reduced order model for three-dimensional unsteady flow using mesh transformation and stitching

Xin Li(李鑫),[1] Zhiwen Deng(邓志文),[2] Rui Feng(冯瑞),[3] Ziyang Liu(刘子扬),[1] Renkun Han(韩仁坤),[1] Hongsheng Liu(刘红升),[2] and Gang Chen(陈刚)[1,a]

[1]*Shaanxi Key Laboratory of Environment and Control for Flight Vehicle, State Key Laboratory for Strength and Vibration of Mechanical Structures, School of Aerospace Engineering, Xi'an Jiaotong University, Xi'an 710049, China*

[2]*Huawei Technologies Co. Ltd, Shenzhen 518129, China*

[3]*Beijing Institute of Space Mechanics & Electricity, Beijing 100094, China*

a) *Author to whom correspondence should be addressed: aachengang@xjtu.edu.cn*

**ABSTRACT**

Artificial intelligence-based three-dimensional(3D) fluid modeling has gained significant attention in recent years. However, the accuracy of such models is often limited by the processing of irregular flow data. In order to bolster the credibility of near-wall flow prediction, this paper presents a deep learning-based reduced order model for three-dimensional unsteady flow using the transformation and stitching of multi-block structured meshes. To begin with, full-order flow data is provided by numerical simulations that rely on multi-block structured meshes. A mesh transformation technique is applied to convert each structured grid with data into a corresponding uniform and orthogonal grid, which is subsequently stitched and filled. The resulting snapshots in the transformed domain contain accurate flow information for multiple meshes and can be directly fed into a structured neural network without requiring any interpolation operation. Subsequently, a network model based on a fully convolutional neural network is constructed to predict flow dynamics accurately. To validate the strategy's feasibility, the flow around a sphere with Re=300 was investigated, and the results obtained using traditional Cartesian interpolation were used as the baseline for comparison. All the results demonstrate the preservation and accurate prediction of flow details near the wall, with the pressure correlation coefficient on the wall achieving an impressive value of 0.9985. Moreover, the periodic behavior of flow fields can be faithfully predicted during long-term inference.

## I. INTRODUCTION

The rapid acquisition and accurate analysis of three-dimensional flow fields have always been a critical and challenging research topic in engineering applications of fluid mechanics. Practical engineering problems typically involve intricate three-dimensional flows that exhibit diverse flow phenomena. Within the confines of the current theoretical framework, the Navier-Stokes equations, which govern these intricate spatiotemporal dynamics, often lack definitive analytical solutions. Over the past few decades, computational fluid dynamics (CFD) has been widely utilized to demonstrate its efficacy and prowess in resolving fluidic predicaments. As the number of grid degrees of freedom in three-dimensional space increases exponentially, analyzing flow fields becomes the most time-consuming aspect of aerodynamic design and optimization processes, necessitating substantial computational resources. The ever-shortening engineering design cycle imposes greater demands on the efficiency and accuracy of flow field simulations.

To address the challenge of "curse of dimensionality" in high-fidelity flow simulations, reduced-order model (ROM)[1] has been proposed to render solutions computationally feasible in real time. Data-driven dimensionality reduction methods can be broadly classified into linear methods, such as subspace projection, and nonlinear methods, including deep learning approaches. Physically interpretable low-dimensional manifolds could be obtained from high-dimensional flow field datasets through techniques such as proper orthogonal decomposition (POD),[2] and dynamic mode decomposition (DMD).[3] However, these linear methods are incapable of capturing the nonlinear nature of multi-scale complex flows, leading to a notable increase in the number of low-order modes. Furthermore, additional governing equations or procedures on the manifold need to be established in order to learn the temporal characteristics of unsteady flow.

Nowadays, deep learning, with its powerful nonlinear mapping capabilities, has emerged as the predominant approach for reduced-order modeling of unsteady flows.[4] Deep learning-based reduced-order model (DLROM) can automatically extract flow rules and multi-scale features from massive flow data and has the flexibility to combine with linear reduction methods. The "black box" neural network model, specifically variants of Convolutional Neural Networks (CNN) with the unique characteristics of local connection and weight sharing,[5] is trained to directly approximate the complete Navier-Stokes equation. CNN was initially utilized by Guo et al.[6] to rapidly estimate non-uniform steady-state velocity fields in both 2D and 3D around blunt bodies, with results indicating a significant reduction in computational costs when compared to CFD. Since then, numerous steady flow models based on CNN have been proposed, demonstrating the capability of CNN in extracting low-dimensional manifolds. For unsteady flow modeling, the time-varying characteristics are primarily inferred through the utilization of recurrent neural networks such as long short-term memory networks and gated neural networks based on a large dataset consisting of numerous historical manifolds obtained by CNN. It is worth noting that deep learning-based flow modeling[7-14] mainly focuses on 2D shapes and encounters more difficult challenges when extended to 3D. This is primarily due to the increased presence of spatial coupling effects and time-series features in 3D flow data with higher dimensions compared to their 2D or 2.5D counterparts. The extraction of nonlinear features necessitates the establishment of more intricate neural network architectures and appropriate learning algorithms. Simultaneously, the processing of extensive data sets and the training of neural networks with a substantial number of model parameters necessitate robust computational capabilities and storage resources,[15,16] which may be constrained by hardware limitations. Recently, advanced techniques such as transfer learning have been applied to reduce the training burden of 3D reduced-order models.

Although CNN is an effective and flexible method for dimensionality reduction, it is necessary to preprocess the 3D flow field into an image using Cartesian uniform interpolation in order to perform discrete convolution. When dealing with flow fields that feature complex boundaries, geometric information is typically represented in one of two ways: a signed distance function[6] or a binary mask.[17] Due to the uniformity of the interpolation grid, there may be a partial loss of flow structure and geometric information,[18] particularly in the 3D near-wall region. Essentially, the uniform scale does not align with the non-uniform multi-scale nature of partial differential equations, leading to a decrease in prediction accuracy for downstream tasks.[19] However, accurately obtaining flow information on the geometric surface, which is of particular concern in engineering design and other fields, remains a challenging task.

One possible solution is to integrate the discretization scheme in CFD with DLROM. The discretized grid, which incorporates flow inhomogeneity and geometric adaptability, naturally supplies the model with multi-scale information. The graph neural network, which possesses a stronger inductive bias, overcomes the limitation of Euclidean space and offers a modeling solution for unstructured grids.[20] However, the current graph neural network still has limitations when it comes to representing dynamic graphs, message passing settings, and other factors, especially in the context of large-scale graphs. Indeed, we have redirected our focus towards structured grids, and their compatibility with well-established and efficient convolutional neural networks has proven to be highly satisfactory. The widely utilized structured grid is also non-uniform, necessitating the implementation of coordinate transformation technology to acquire a uniform mesh in the computational space that can be integrated with advanced structured neural networks. Coordinate transformation technology establishes a one-to-one mapping between physical space coordinates and computational space coordinates. In other words, the reduced-order flow model, viewed from the perspective of the transformation domain, replaces the computational fluid dynamics (CFD) calculation in the computational space. Consequently, the prediction results also require coordinate inverse transformation to obtain the actual physical flow. Chen *et al.*[21] were the first to utilize coordinate transformation to establish an independent reduced-order model for predicting the steady-state flow field of an airfoil. They also investigated the impact of various forms of coordinate transformation on deep learning. The results demonstrate that the method achieves a reduction in error by an order of magnitude compared to Cartesian interpolation. This provides substantial evidence supporting the effectiveness and practicality of the employed strategy. Subsequently, Deng *et al.*[22] successfully employed grid interpolation and transformation techniques to preserve flow details near the boundary in the flow around the cylinder and the transonic buffeting flow field. They combined these techniques with two advanced neural networks, namely Unet and FNO, and successfully applied them to unsteady periodic flows. Recently, Hu *et al.*[23] achieved high-precision prediction of the subsonic steady flow field of a three-dimensional airfoil with millions of grid nodes from the transform domain. In summary, flow modeling from the transformed domain perspective has shown significant potential.

However, current researches primarily focus on single-block structured grid transformations, particularly for modeling steady flow fields. As shape complexity increases, a single mesh becomes inadequate for partitioning the solution domain. In practice, the solution domain is often divided into several regions with simpler shapes, and each region is then separately generated with a structured mesh, forming a multi-block structured mesh. This study emphasizes the deep integration of multi-block structured grids with convolutional neural networks. To the best of our knowledge, this paper represents the first attempt at applying this approach to three-dimensional unsteady flow. This learning strategy enables accurate prediction of flow fields near complex walls without the need for additional recovery procedures. Furthermore, we have introduced a gradient term with weak physical interpretability to the loss function, which provides additional constraints to enhance the model's performance. To evaluate the feasibility of our proposed strategy, we conducted experiments on a three-dimensional incompressible flow around a sphere with Reynolds number equal to 300.

The manuscript is organized as follows: In Sec. II, we provide a brief introduction to the governing equations for full-order simulations and the mesh transformation and stitching method, comparing it with the uniform interpolation method; Sec. III covers the basic structure, loss function, and specific training strategy of the neural network; Furthermore, Sec. IV presents the flow configuration around a sphere, providing a detailed comparison of prediction results based on different methods and evaluating the long-term inference performance of the neural networks; Finally, conclusions are provided in Sec. V.

## II. METHODOLOGY

Data-driven reduced-order modeling relies on full-order flow field snapshots as the ground truth for future training. In this section, we will begin by describing the full-order simulation process, specifically based on multi-block structured meshes. Subsequently, we will delve into the mesh transformation and stitching method, providing a comparison with the uniform interpolation method.

## A. Full-Order Modeling Based on Multi-Block Meshes

The rich and complex dynamic behavior of fluids can be described by the Navier-Stokes (NS) equations. In the Cartesian coordinate system, the NS governing equations for incompressible viscous fluids consist of the following continuity equation and the momentum equation,

$$\nabla \cdot u = 0, \tag{1}$$

$$\frac{\partial u}{\partial t} + (u \cdot \nabla)u = -\frac{1}{\rho}\nabla p + \nu \nabla^2 u, \tag{2}$$

Where $\nabla = (\partial/\partial x, \partial/\partial y, \partial/\partial z)$ denotes the Hamilton operator, $u = (u, v, w)$ represents the fluid velocity in the respective Cartesian directions, and $p$, $t$, $\rho$, and $\upsilon$ are pressure, time, fluid density, and kinematic viscosity, respectively. The Reynolds number is defined as

$$\text{Re} = \frac{U_0 D}{\upsilon}, \tag{3}$$

Here, the free stream velocity $U_0$ and the kinematic viscosity $\upsilon$ are adjusted by *Re*.

To integrate deep learning, a multi-block structured grid is employed to discretize the fluid domain, striking a balance between calculation accuracy and cost. Each block that makes up the overall mesh is independent of the other and has a corresponding interface. The commercial CFD code Fluent 2021R2, based on the finite volume method, is utilized to solve the governing equations in a discrete Euler field. The simulation is performed in transient mode using the segregated solver, the SIMPLE Pressure-Velocity coupling, second-order upwind discretization, and the cell-based gradient option. A first-order implicit scheme is applied to discretize the time term. The calculation process typically involves iteratively running with smaller time steps until convergence is attained.

**B. Cartesian Uniform Interpolation Method**

In previous studies on deep learning reduction models, full-order flow field snapshots were commonly transformed into regular data using the Cartesian uniform interpolation (CUI) method before being fed into neural networks. The drawback of using a uniform grid is its inability to allocate additional nodes specifically in critical areas of the flow. Instead, it can only uniformly increase the number of nodes across the entire grid, resulting in a significant increase in memory requirements, particularly in 3D cases. For the sake of clear comparison, this paper employs the Cartesian uniform interpolation (CUI) method as the baseline method.

The specific implementation process of the CUI method is as follows:

(a) First, a cuboid space, known as the sampling space, is selected as the designated deep learning space $\Omega^1$. Uniformly distributed voxel probes with a resolution of $N_x \times N_y \times N_z$ are placed throughout the space, with the numbers indicating the number of probes in different Cartesian directions.

(b) Then, Inhomogeneous information values within the flow field, akin to pixel values in an image, are projected onto the uniformly distributed voxel probes using linear interpolation. It has been proven that linear interpolation, which avoids flow field oscillation, exhibits good linear convergence as the resolution increases.

(c) Additionally, drawing inspiration from binary masks, variables within geometric regions where there is no flow are typically assigned constant values. Another effective approach involves constructing a distance sign function to identify boundary information, but this may introduce unnecessary information and increase the training burden.

(d) A total of *C* flow characteristic variables, such as pressure (p), velocity (u), etc., were meticulously selected for research and organized into a 3D flow snapshot matrix $\mathbf{F} = \{\mathbf{F}^1, \mathbf{F}^2, ..., \mathbf{F}^T\} \in \mathbb{R}^{T \times C \times N_x \times N_y \times N_z}$, where the first 3D flow snapshot is $\mathbf{F}^1 \in \mathbb{R}^{C \times N_x \times N_y \times N_z}$, and *T* is the time steps.

## C. Mesh Transformation and Stitching Method

The information loss resulting from uniform interpolation is deemed unacceptable in engineering fields such as strength design. As a result, there is a pursuit of combining structured grids with deep learning to enhance flow accuracy.

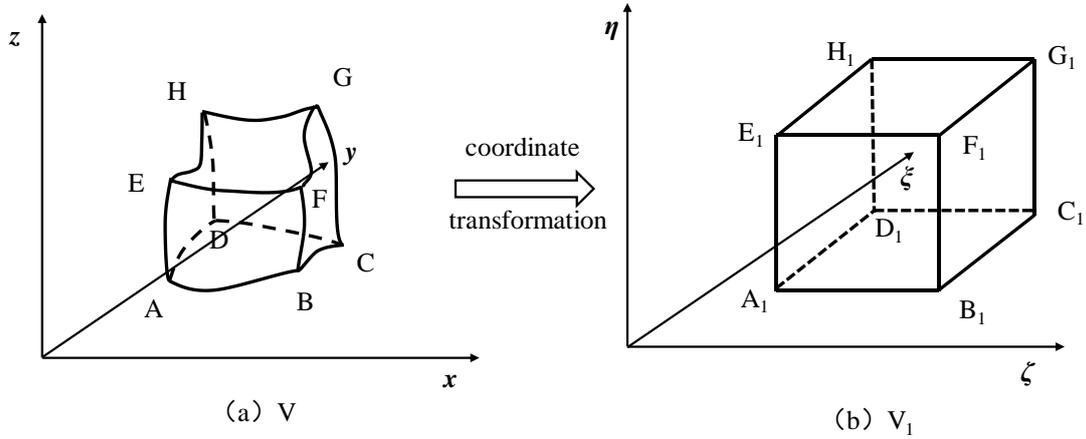

Fig. 1. Schematic diagram of coordinate transformation of a single region in (a) physical space V and (b) computational space $V_1$.

Mesh transformation, which has been employed in computational fluid dynamics (CFD) for a considerable period, plays a pivotal role in the integration of structured grids and structured networks. As depicted in Fig. 1, there is an irregularly shaped object domain $V$ in the physical coordinate system (a) *xyz*. However, through a specific coordinate transformation, we can convert it into a regular cube $V_1$ in the transformed space (b) $\zeta\eta\xi$. For the given structured grid, there exists a one-to-one correspondence between the mesh points in the two coordinate systems, as follows:

$$\begin{cases} \xi = \xi(x, y, z) \\ \eta = \eta(x, y, z) \\ \zeta = \zeta(x, y, z) \end{cases}, \tag{4}$$

Then, we have

$$\begin{bmatrix} d\xi \\ d\eta \\ d\zeta \end{bmatrix} = \begin{bmatrix} \partial_x\xi & \partial_y\xi & \partial_z\xi \\ \partial_x\eta & \partial_y\eta & \partial_z\eta \\ \partial_x\zeta & \partial_y\zeta & \partial_z\zeta \end{bmatrix} \begin{bmatrix} dx \\ dy \\ dz \end{bmatrix}, \tag{5}$$

which is a univalent transformation. The definition of $\xi$, $\eta$, and $\zeta$ with given integers $i$ from 1 to $i_{max}$, $j$ from 1 to $j_{max}$, and $k$ from 1 to $k_{max}$ is

$$\begin{cases} \xi = \dfrac{i-1}{i_{max}-1} \in [0,1] \\ \eta = \dfrac{j-1}{j_{max}-1} \in [0,1] \\ \zeta = \dfrac{k-1}{k_{max}-1} \in [0,1] \end{cases}, \tag{6}$$

where $i$, $j$, and $k$ denotes the indices of the uneven mesh in different directions. Up to this point, the transformation matrix can be constructed to realize the unique mapping of a single structured grid.

Meshes in the transform domain fulfill the criteria of orthogonality and regularization, allowing them to be directly employed as inputs to structured neural networks. However, in practical 3D engineering applications, multi-block structured grids are typically employed as part of the discretization scheme, as described in the present paper. While it is feasible to input each structured mesh independently as an isolated entity, this modeling approach disregards the interaction of the flow field on the shared interface.

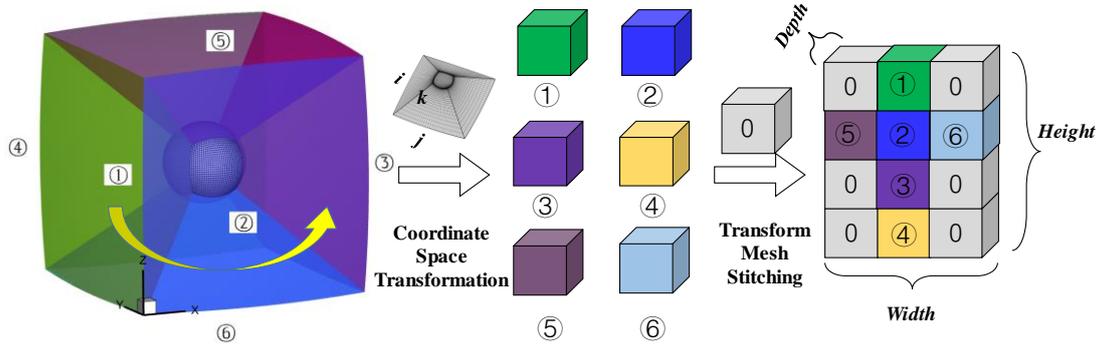

Fig. 2. Schematic diagram of the transformation, stitching, and filling of multi-block structured grids

We begin by conducting mesh transformations for multi-block meshes, followed by the implementation of specifically designed mesh stitching and padding techniques. We have embarked on a groundbreaking endeavor to stitch the transformed regular cube data. The principle behind this stitching process is to preserve the topological connectivity of the physical space to the greatest extent possible. As illustrated in Fig. 2, the region surrounding the sphere is subdivided into six irregular areas, marked with numbers ①-⑥, each of which is further divided into structured meshes. Each structured grid has the same resolution $i_{max} \times j_{max} \times k_{max}$. Initially, each structured mesh undergoes coordinate transformation to produce a regular cuboid mesh with dimensions of $i_{max} \times j_{max} \times k_{max}$. These transformed meshes are then stitched together, based on the corresponding relationships of their connecting interfaces, with the goal of maximizing the continuity of physical quantities within the connection regions. The blank regions resulting from the stitching process are filled with a constant-value gray cubic block, thereby creating a larger and more regular rectangular cuboid mesh with dimensions of

$H \times W \times D$. There are $C$ flow variables that need to be transformed, which are consistent with the selection in the CUI method. Ultimately, we obtained a similar dataset $\mathbf{F} = \{\mathbf{F}^1, \mathbf{F}^2, ..., \mathbf{F}^T\} \in \mathbb{R}^{T \times C \times H \times W \times D}$ of three-dimensional flow snapshots in the transformed domain.

To keep things concise, the mesh transformation and stitching method described above is referred to as MTS in the present study. It is important to note that the flow snapshots in the transformed domain, inferred by the neural network, need to be remapped back to the actual physical space using the inverse transformation. In other words, the learning task of the neural network model is shifted to the transform domain.

**D. Data Preprocessing Pipeline**

To enhance the convergence speed and performance of the model, a unified pipeline for data preprocessing is established to address scale differences prior to formal training. Initially, a non-dimensionalization preprocessing step with physical interpretability is employed to ensure that the learned physical laws are independent of dimensional scales. The channel information within the flow field snapshot, consisting of pressure and velocity variables, is then processed using different non-dimensionalization scales, as outlined below:

$$p^* = \frac{p}{\rho U_0^2}, \tag{7}$$

$$u^* = \frac{u}{U_0}. \tag{8}$$

Next, we apply the time-averaged flow field and min-max normalization methods to centralize the entire dataset $\mathbf{F} \in \mathbb{R}^{T \times C \times H \times W \times D}$.[24] The time-averaged field $\bar{\mathbf{F}} \in \mathbb{R}^{C \times H \times W \times D}$ and fluctuating flow fields $\mathbf{F}' \in \mathbb{R}^{T \times C \times H \times W \times D}$ over the entire dataset are calculated as follows,

$$\bar{\mathbf{F}} = \frac{1}{T}\sum_{i=1}^{T}\mathbf{F}^i, \tag{9}$$

$$\mathbf{F}'^i = \mathbf{F}^i - \bar{\mathbf{F}}. \tag{10}$$

Then the max-min normalization method is employed to scale the variables $\mathbf{F}'$ to the range [0,1],

$$\mathbf{F}'^i_s = \frac{\mathbf{F}'^i - \min}{\max - \min}, \tag{11}$$

where $\min \in \mathbb{R}^C$ and $\max \in \mathbb{R}^C$ are the maximum and the minimum values in $\mathbf{F}'$, respectively. Finally, the dataset, $\mathbf{F}'_s = \{\mathbf{F}'^1_s, \mathbf{F}'^2_s, \cdots, \mathbf{F}'^T_s\} \in \mathbb{R}^{T \times C \times H \times W \times D}$, is divided into a training set with 0.7T time steps, a test set with 0.1T time steps, and an inference set with 0.2T consecutive time steps.

## III. DEEP LEARNING-BASED REDUCED ORDER MODEL

In this section, we propose a reduced-order model based on a fully convolutional neural network for predicting three-dimensional unsteady flow. Acquiring a significant number of high-precision three-dimensional unsteady flow snapshots is a challenge in practical engineering design. Unlike the approach of analyzing a large amount of historical time-step data using recurrent neural networks, the objective of this study is to directly establish an estimation $f$ of future flow fields $\hat{\mathbf{F}}^{i+1}$ based on the current flow snapshot $\mathbf{F}^i$,

$$\hat{\mathbf{F}}^{i+1} = f(\mathbf{F}^i, \theta), \tag{12}$$

where $f$ is a neural network model parameterized by $\theta$. Furthermore, the iterative prediction strategy is adopted to simulate the complete evolution process of the unsteady flow.

## A. Architecture of Neural Network

We employ 3D Unet[25] and ResNet[26] architectures to construct the deep neural network model, ResUnet3D, as illustrated in Fig. 3. The 3D U-Net is a fully convolutional neural network designed for processing three-dimensional input images, making it well-suited for extracting features from three-dimensional flow snapshots. While most studies focus on developing reduced-order models for steady flow fields, they often overlook the model's capability to capture temporal evolution. Generally, as the depth of a convolutional neural network increases, its nonlinear expressive power strengthens. However, increasing the number of layers can also lead to accuracy degradation. To address this, ResNet employs residual connections to mitigate the performance degradation associated with deep networks, making it a reliable backbone for extracting engineering data features.[27]

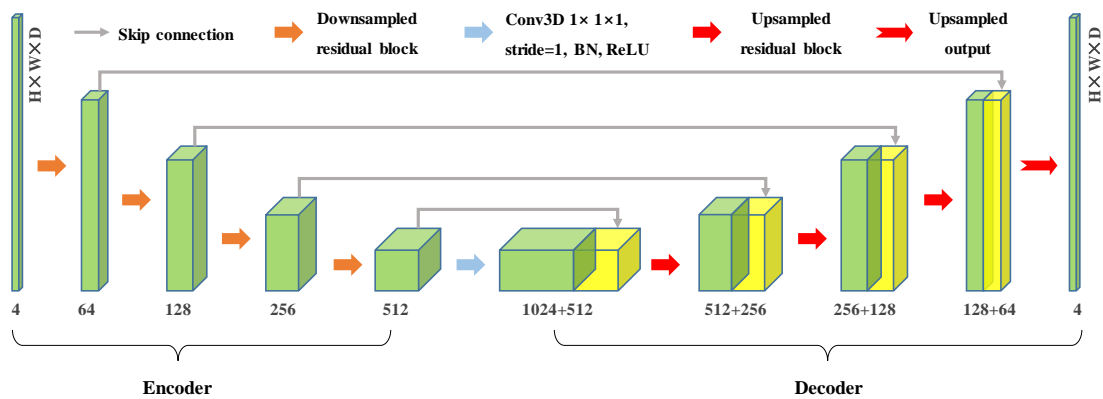

Fig. 3. The architecture of the deep neural network ResUnet3D.

The proposed neural network model follows the paradigm of an encoder-decoder architecture, which exhibits a symmetrical U-shaped structure. The main difference lies in the replacement of traditional convolutions with convolutional residual blocks.

**Encoder**: The left side of the network, known as the contracting path, is responsible for hierarchically extracting the latent features of the high-dimensional flow field. The encoder consists of four downsampled residual blocks, as illustrated in Fig. 4(a). Residual connections can be represented as

$$Y = F(X, \{\mathbf{w}_i\}, \sigma) + H(\mathbf{x}, \mathbf{w}_j), \tag{13}$$

where $F$ consists of two convolution operations with the LeakyReLU activation function. Following each convolution operation, batch normalization is applied to ensure a more stable distribution of the parameter $\mathbf{w}$. $H$ represents the convolution operation without an activation function, but it is crucial that the output has the same characteristic dimension as $F$. Downsampling is accomplished by utilizing convolutional operations with a stride of 2 instead of pooling operations. Following each residual block operation, the number of feature channels is doubled while the size of the feature map is halved.

**Decoder**: On the right side of the network, referred to as the expansive path, the low-dimensional features are upsampled. Correspondingly, the decoding part also includes four upsampling residual blocks, with the structure of the upsampling residual block shown in Fig. 4 (b). The first step involves the application of deconvolution to increase the size of the original features by a factor of two while reducing the number of feature channels. Next, a residual connection structure is formed by combining convolutional operations with an identity connection. Inspired by previous work[26], in addition to the residual connections within the residual blocks, we also introduced skip connections in our model, indicated by solid gray arrows in Fig. 3. The increased number of residual connections helps in capturing low-frequency features of the high-dimensional flow field, further enriching the details of the flow field prediction. It is essential to note that the structure of the upsampled output block, which omits the identity connection $H$, is responsible for the final output of the model.

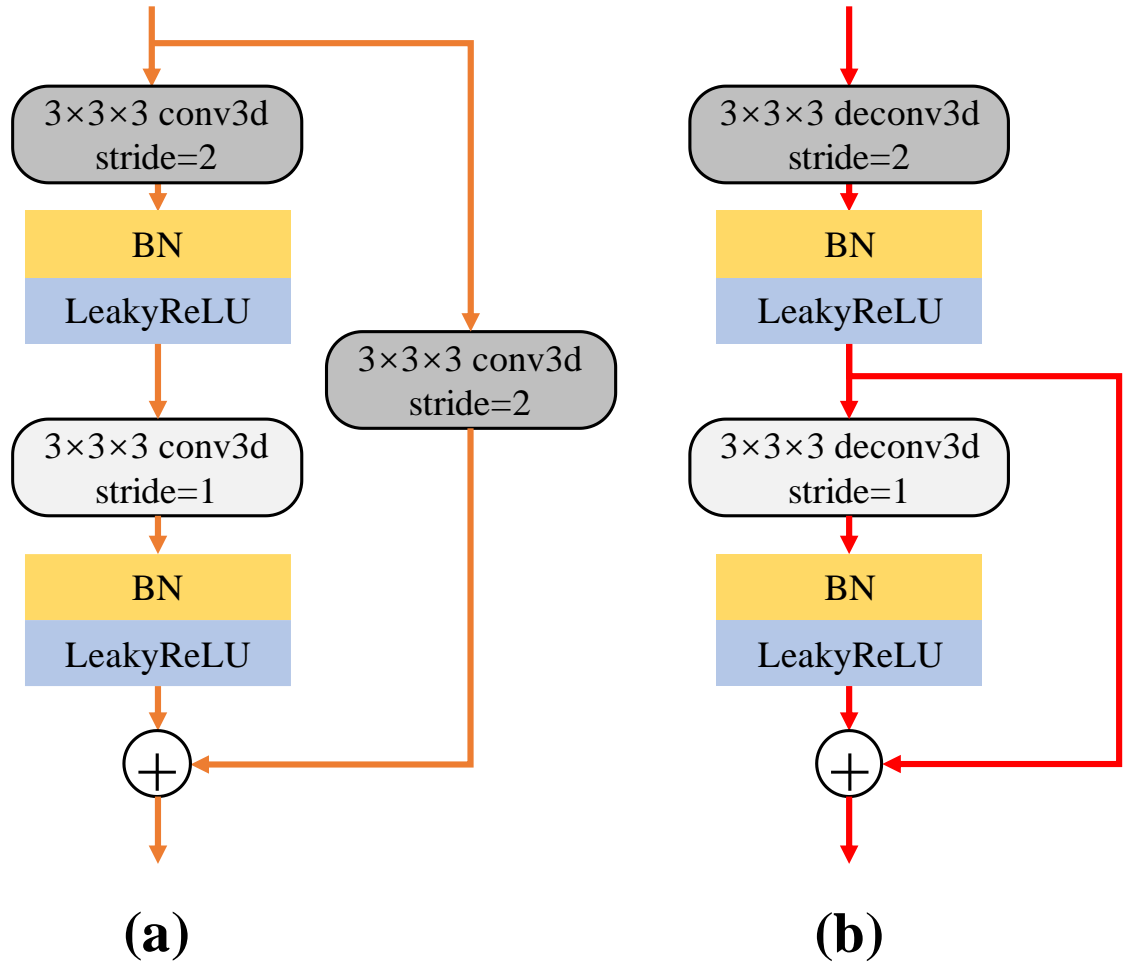

Fig. 4. Schematic diagram of residual block: (a) a downsampled residual block and (b) an upsampled residual block

## B. Loss Function Design

The loss function is a crucial component and serves as the optimization objective during the training process of neural networks. In typical regression problems, Root

Mean Square Error (RMSE) is commonly employed as the loss function. However, considering the varying scales of different variables in the flow, we choose the relative L2 error of different channels as an intensity penalty. This ensures the similarity of all voxels in the channel space,

$$L_{\text{int}} = \sum_{j=1}^{4} \frac{\left\| \mathbf{C}^j - \hat{\mathbf{C}}^j \right\|}{\left\| \mathbf{C}^j - \overline{\mathbf{C}}^j \right\|}, \tag{14}$$

where $\mathbf{C}^j = \mathbf{F}_s'^{ij} \in \mathbb{R}^{H \times W \times D}$ denotes the snapshot of the $jth$ channel of the flow field $\mathbf{F}_s'$ at the time step $i$, $\hat{\mathbf{C}}^j$ represents the corresponding predicted snapshot, and $\overline{\mathbf{C}}^j \in \mathbb{R}$ represents the average value of the current snapshot $\mathbf{C}^j$.

Furthermore, drawing inspiration from the finite difference method in the transformed domain, we introduce a gradient loss function that possesses a certain level of physical interpretability. This gradient loss function helps enhance the sharpness of the generated snapshots,

$$L_{gd} = \sum_{j=1}^{4} \left( \frac{\left\| \mathbf{C}_h^j - \hat{\mathbf{C}}_h^j \right\|}{\left\| \mathbf{C}_h^j - \overline{\mathbf{C}}_h^j \right\|} + \frac{\left\| \mathbf{C}_w^j - \hat{\mathbf{C}}_w^j \right\|}{\left\| \mathbf{C}_w^j - \overline{\mathbf{C}}_w^j \right\|} + \frac{\left\| \mathbf{C}_d^j - \hat{\mathbf{C}}_d^j \right\|}{\left\| \mathbf{C}_d^j - \overline{\mathbf{C}}_d^j \right\|} \right), \tag{15}$$

where, $\mathbf{C}_h^j = \frac{\partial}{\partial h} \mathbf{C}^j, \mathbf{C}_w^j = \frac{\partial}{\partial w} \mathbf{C}^j$, and $\mathbf{C}_d^j = \frac{\partial}{\partial d} \mathbf{C}^j$ represent the first-order derivatives of the flow field variables along the $h$, $w$, and $d$ directions, respectively. Gradients on a regular grid of points are approximated using a second-order central difference scheme. The mixed loss used in this paper is as follows,

$$Loss = \alpha \cdot L_{\text{int}} + \beta \cdot L_{gd} \tag{16}$$

where $\alpha$ and $\beta$ are the coefficients that determine the contribution of the respective loss term.

### C. Training Strategy

The neural network model in this study was constructed using the latest MindSpore[28] framework, which is a self-developed and open-source framework by Huawei. For model training, the Adam optimization algorithm with gradient weight adaptation was employed. Compared to other optimization algorithms, the Adam optimizer offers advantages such as high computational efficiency and low memory consumption. The learning rate was adjusted using the cosine annealing strategy, starting with an initial value of 0.01.

In addition, the adjustment of the loss function weights relies on the adaptive adjustment strategy implemented in the MindFlow code library for multitask learning. The training was conducted for 500 epochs, with a batch size of 10. Each training session on the Nvidia GeForce RTX 3090 24M took approximately 3.5 hours to complete.

## IV. RESULTS AND DISCUSSION

In this section, we evaluate the MTS method and the data-driven ResUnet3D model using a representative flow around a sphere with Re=300. Our focus is on assessing the accuracy of flow prediction near the wall, particularly highlighting the improvement achieved through the stitching strategy compared to CUI. Additionally, we examine the flow feature capture capability of the neural networks, with specific attention given to the temporal evolution relationship of physical quantities within the flow field over time.

## A. Flow Configuration

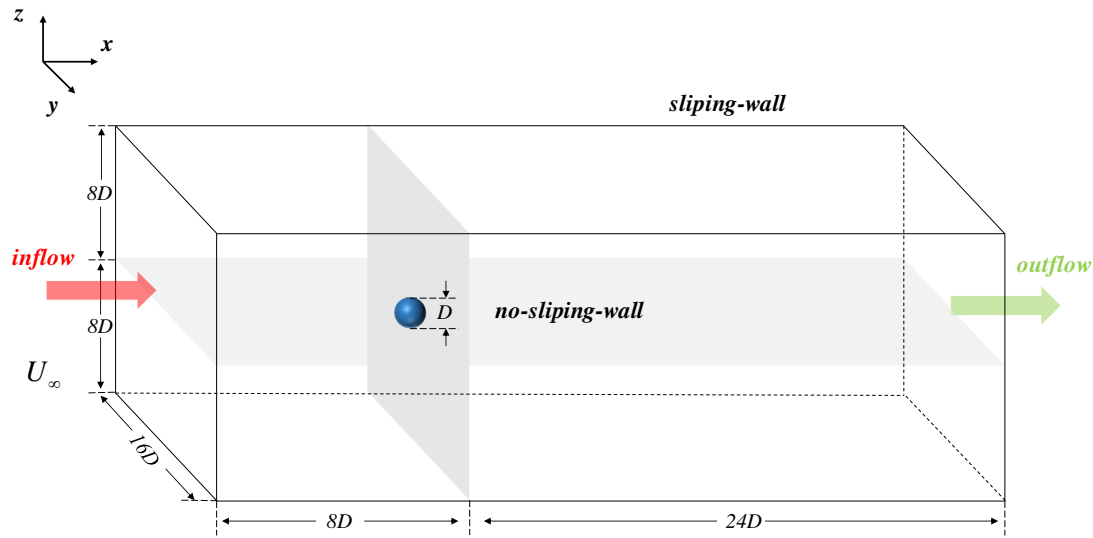

Fig. 5. Flow configuration information for flow around a sphere, including geometry and boundary conditions.

The schematic diagram of the flow configuration related to the flow around the sphere is shown in Fig. 5. A sphere of diameter $D$=1m is located at the origin of the Cartesian coordinate system. The outer far-field domain is a cuboid with dimensions 32$D$, 16$D$, and 16$D$ in the $x$, $y$, and $z$ directions, respectively. The distance from the center of the sphere to the left boundary and the circumferential wall is set to 8$D$, which is considered large enough to ensure that the boundary conditions do not interfere with the flow of the sphere. A uniform incoming flow boundary condition is applied to the inlet, with the flow direction along the x direction. In addition, a pressure outflow boundary condition is imposed on the right outlet; a no-slip boundary condition is maintained on the spherical wall, while the slip boundary condition is implemented on the remaining boundaries.

The block strategy and grid division results adopted in the whole domain are shown in Fig. 6, and the total number of hexahedral grids reaches 218w. In particular, special attention has been paid to the near-wall region $\Omega$ of dimension $3D \times 3D \times 3D$ centered on the sphere, which is divided by a 3D O-shaped mesh. Each grid block within this area has a size of $i_{max} \times j_{max} \times k_{max} = 32 \times 32 \times 32$. The first cell height is set to about 15 mm (for $y^+<1$). A simulation time of 100 seconds is large enough to guarantee the occurrence of any unstable solutions. The simulation time $t=100s$ with a time step $\Delta T = 0.0125s$ resulted in the appearance of a regular periodic solution, which took about 50 hours on a CentOS system with 32 processors.

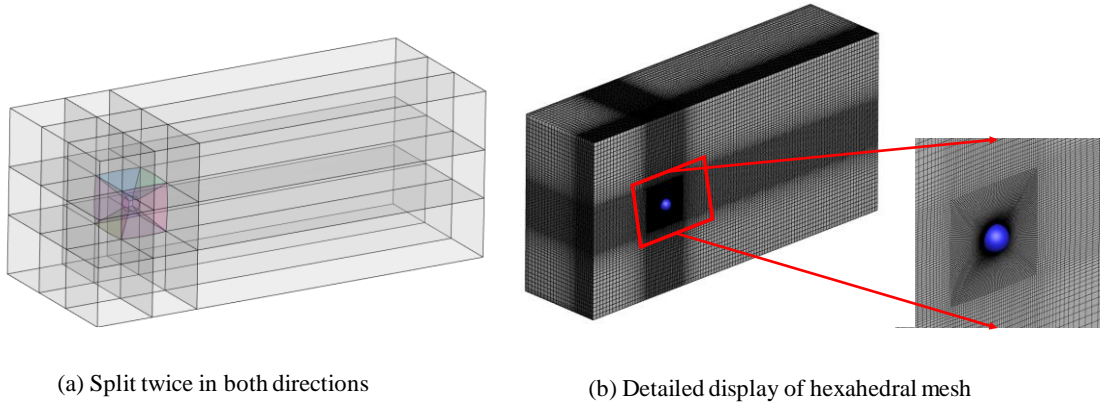

(a) Split twice in both directions    (b) Detailed display of hexahedral mesh

Fig. 6. Blocking strategy and meshing results for the entire domain.

Fig. 7 provides the force coefficients versus time between t = 95s and t = 135 s, the lift coefficient $C_L = F_y / \left( \left( \frac{1}{2} \rho U_\infty^2 \right) \left( \pi D^2 / 4 \right) \right)$, and the drag coefficient $C_D = F_x / \left( \left( \frac{1}{2} \rho U_\infty^2 \right) \left( \pi D^2 / 4 \right) \right)$. Computed average values are $\bar{C}_d = 0.665$ and $\bar{C}_l = 0.0023$, with oscillation amplitudes $\Delta C_D = \left( C_{D_{max}} - C_{D_{min}} \right) / 2 = 2.3 \times 10^{-3}$ and $\Delta C_L = 0.85 \times 10^{-2}$. The results are shown to be in good agreement with previous numerical and experimental work, as shown in Table I. Saving a snapshot of the flow field every 6-time steps after the calculation converges makes about 30-40 snapshots in a flow cycle guaranteed. Finally, a total of 400 snapshots between t=100s and t=130s are saved. The size of the flow field snapshot set obtained by processing the multi-block grids near the sphere using the MTS method, as shown in Fig. 2, is $\mathbf{F} \in \mathbb{R}^{T \times C \times H \times W \times D} = \mathbb{R}^{400 \times 4 \times 128 \times 96 \times 32}$. The corresponding CUI method uniformly places probes in the region $\Omega$, resulting in a dataset with a dimension of $\mathbf{F} \in \mathbb{R}^{T \times C \times N_x \times N_y \times N_z} = \mathbb{R}^{400 \times 4 \times 64 \times 64 \times 64}$.

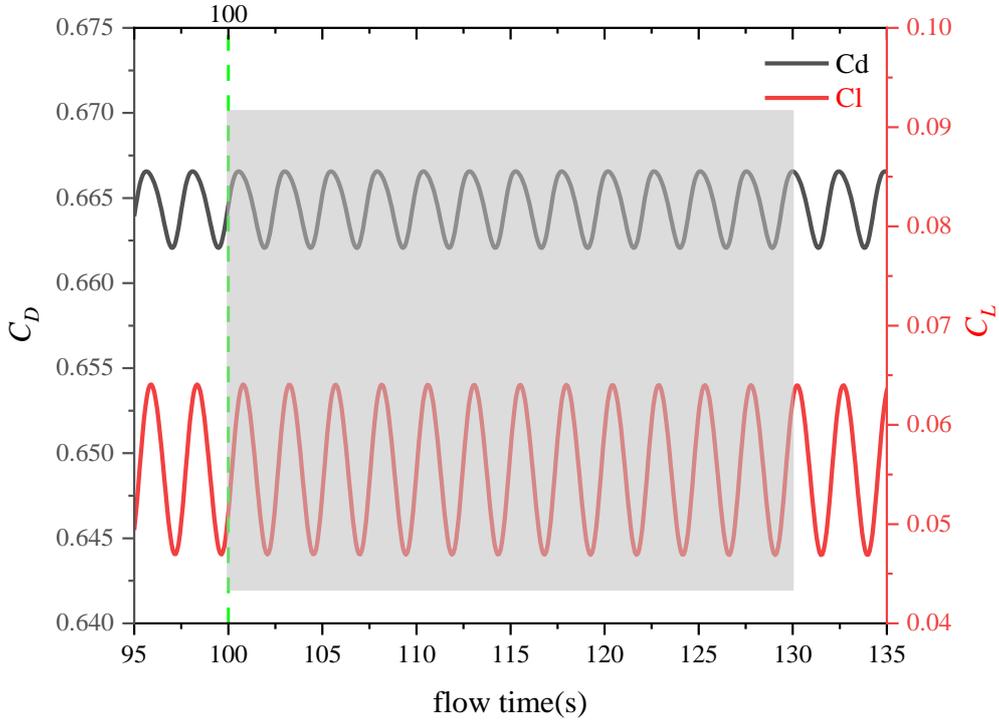

Fig. 7. The change curves of the lift coefficient $C_L$ (red) and the drag coefficient $C_D$ (black).

Table I. The results of the current numerical simulation compared with the listed literature

| Study(Re=300) | $\bar{C}_d$ | $\triangle C_d \times 10^{-3}$ | $\bar{C}_l$ | $\triangle C_l \times 10^{-2}$ | $St$ |
| --- | --- | --- | --- | --- | --- |
| **Present** | 0.665 | 2.3 | 0.056 | 0.85 | 0.135 |
| Ploumhans, et al[29] | 0.683 | 2.5 | 0.061 | 1.4 | 0.135 |
| Johnson and Patel[30] | 0.656 | 3.5 | 0.069 | 1.6 | 0.137 |

## B. Flow Prediction in The Near-Wall Flow Field

To ensure rigor, a hyperparameter search was conducted for the model, and the same set of suitable hyperparameters was determined for training on both datasets, as explained in Section III.C. In order to have consistent evaluation criteria, the L2 relative error distance was utilized as a quantitative metric in the model testing, which aligns with the intensity loss $L_{int}$ as follows:

$$L_2 - Relative\ Error = L_{int} = \sum_{j=1}^{4} \frac{\left\|C^j - \hat{C}^j\right\|}{\left\|C^j - \bar{C}^j\right\|}. \quad (17)$$

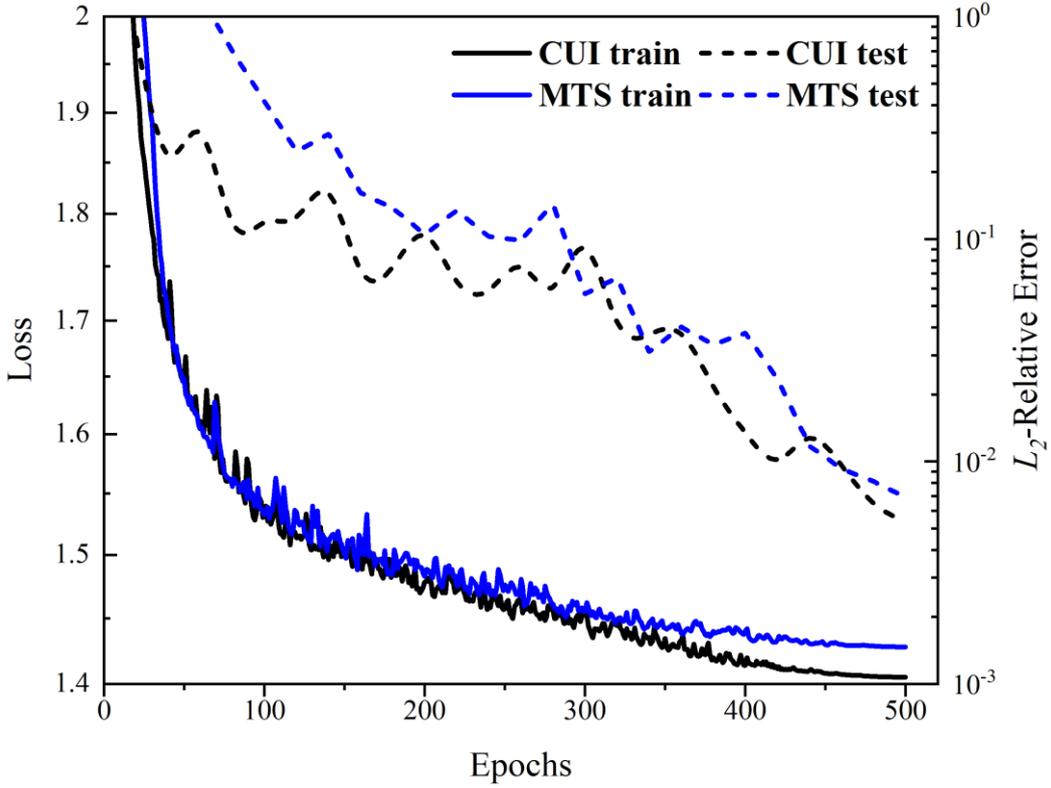

Fig. 8. The training loss and the test relative error of the neural network over epochs on the CUI-based and MTS-based datasets

The changes in training loss and test relative error with training epochs are presented in Fig. 8. The curves are color-coded to differentiate between the different types of datasets, while the line segments represent the training and testing processes. As can be seen from Fig. 8, training loss indicated by the solid line initially decreases rapidly within the first 100 epochs and then gradually converges between epochs 100 and 500.

The CUI-based loss achieves a lower value of 1.405 compared to MTS. As for the relative error curves of the tests, they exhibit fluctuations during the descent but ultimately reach a similar minimum error level of less than $10^{-2}$

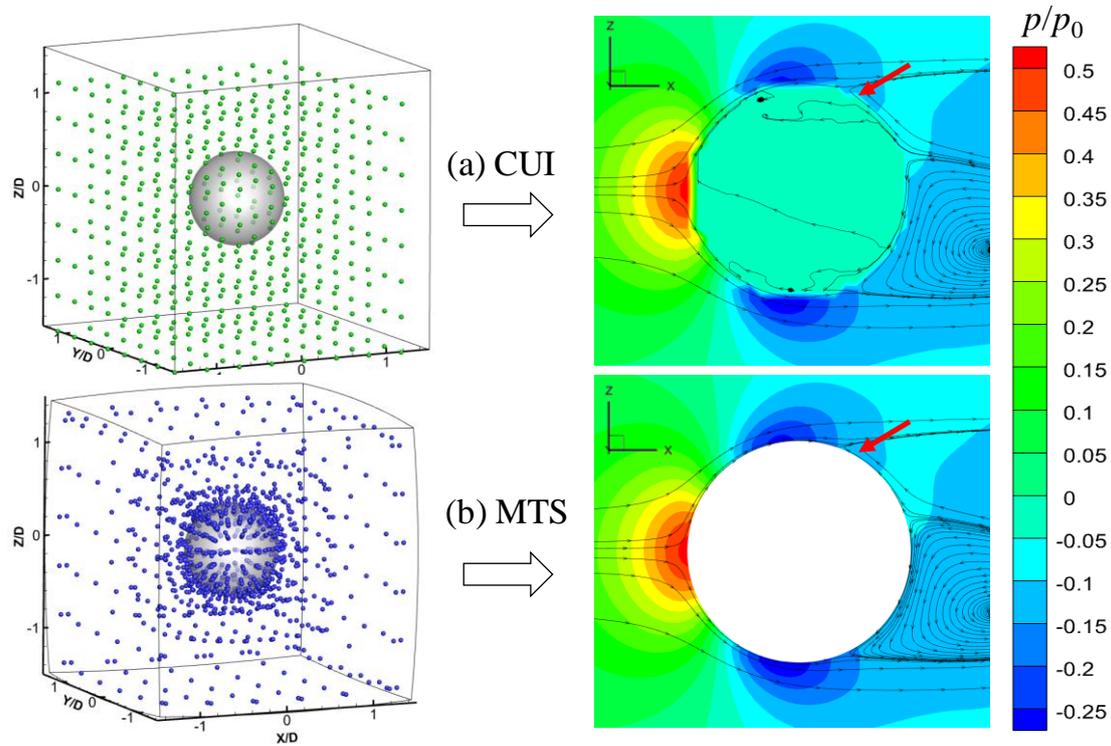

Fig. 9. (a) CUI. (b) MTS. Spatial distribution of voxel probes (left), non-dimensional pressure contour, and the streamline of XZ cross-section (Y=0) predicted at a certain moment (right).

Next, We place particular emphasis on the improvement of the MTS method for predicting the flow field near the wall. One of the test results obtained from the neural network is displayed in the right column of Fig. 9, which includes the non-dimensional pressure contour and streamline distribution of the XZ cross-section (Y=0). Comparing Fig. 9, it is evident that the pressure field predicted using the CUI method near the wall is characterized by an unsmooth distribution and a significant jaggedness. This can be attributed to the uniform distribution of CUI's green probes within the left column of Fig. 9, with some of these probes even strategically positioned within the sphere. In contrast, the MTS-based flow prediction overcomes these shortcomings and can accurately identify the precise location of flow separation. Furthermore, the flow situation on the spherical surface is extracted and compared between the two prediction results, as illustrated in Fig. 10. The CUI method fails to accurately reconstruct the flow on the spherical wall due to insufficient grid resolution near the wall.

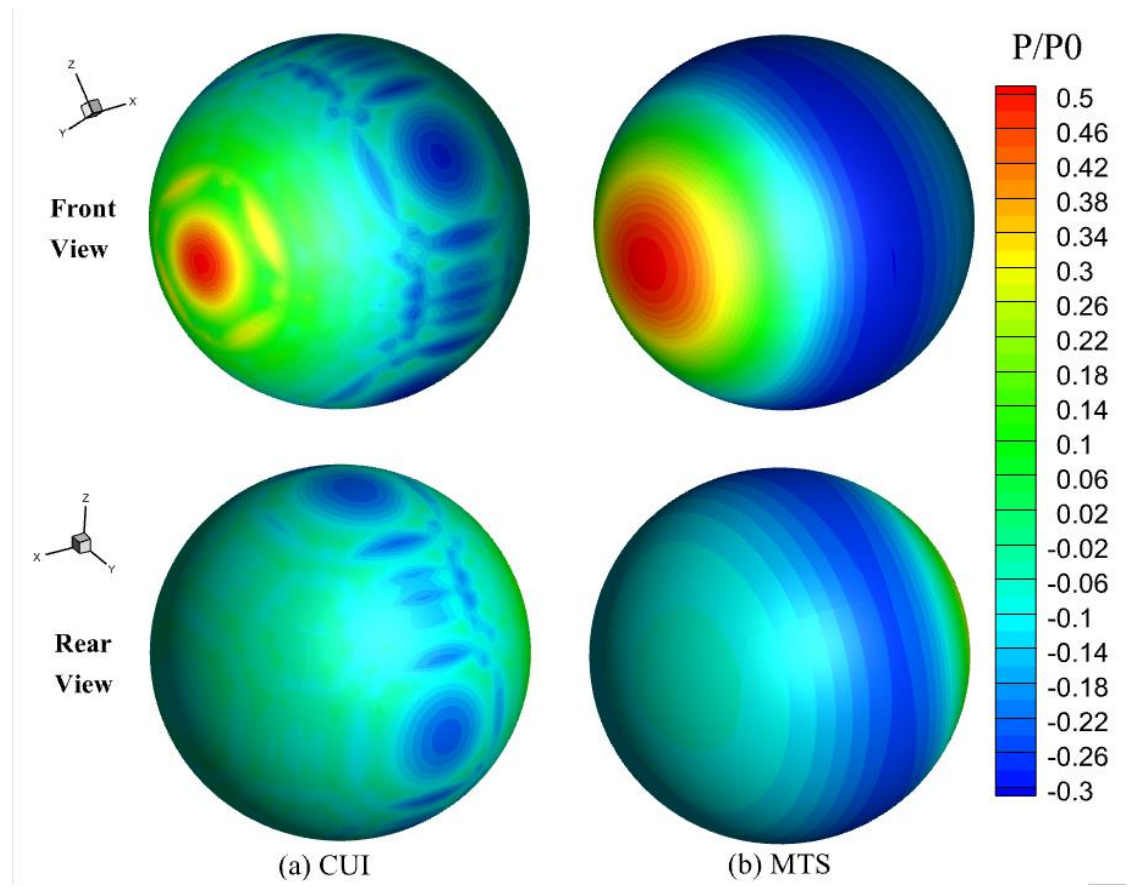

Fig. 10. Flow restoration on the surface of a sphere under different viewing angles. (a)(c) CUI-based results. (b)(d) MTS-based Results.

In addition, we conducted quantitative comparisons of the flow field on the spherical surface by placing measurement points. Two perpendicular directions on the sphere are chosen to place 64 sequential measuring points, respectively, as shown in Fig. 11 (a). The red curve in Fig. 11 (b) represents the result on the red point. It can be observed that the maximum value of the pressure coefficient is predicted similarly across all methods. The prediction trends of all methods are basically consistent with CFD, although there are still jagged fluctuations present in the CUI predictions, which is consistent with the phenomenon we see in Fig. 11 (c). Especially in Fig. 11(c), the prediction coefficient of CUI deviates significantly from the original results, rendering it practically useless for guiding engineering applications. However, we are pleased to note that all the predicted results obtained using the MTS method are consistently aligned with the CFD results. This finding confirms the feasibility and effectiveness of the MTS strategy.

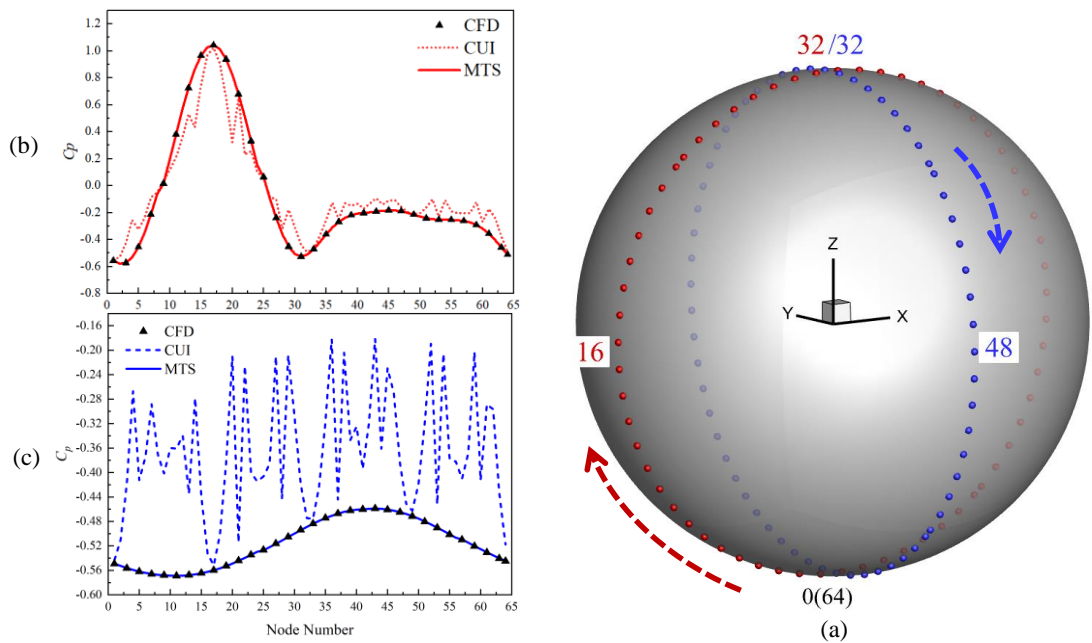

Fig. 11. (a) The numbering sequence of the pressure measurement points on the spherical surface. The plane composed of red points is parallel to the incoming flow, and the plane composed of blue points is perpendicular to the incoming flow. (b) Pressure coefficient values at different red points(c) Pressure coefficient values at different blue points

## C. Evaluation of Neural Network Inference Performance

Various recent test results have demonstrated the model's high accuracy in short-term forecasting. Building upon these findings, the predicted flow field is then redefined as the input for a new round of neural networks. In other words, an iterative forecasting strategy is employed to assess the model's performance in forecasting multiple time steps.

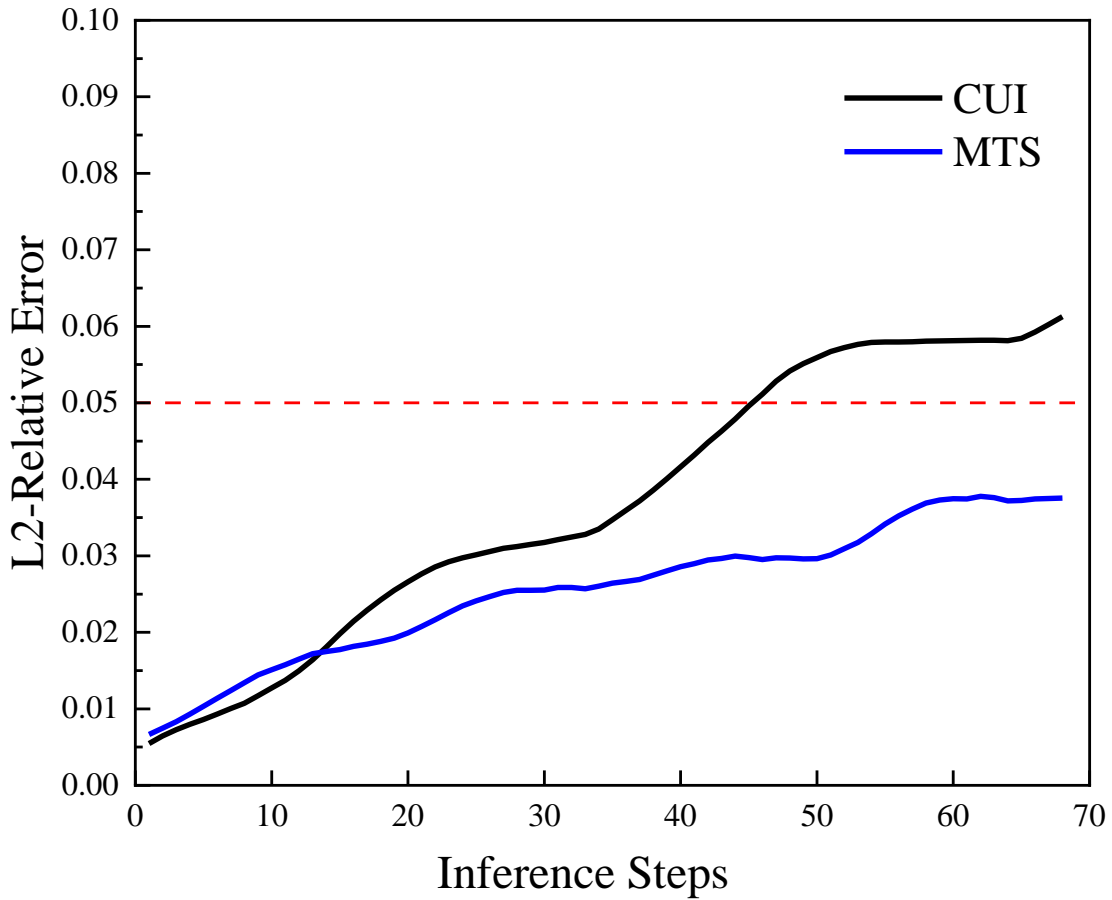

Fig. 12. The L2-relative error curves of CUI-based and MTS-based models when inferring more time steps

The relative error of the model's multi-step inference on the two datasets is plotted in Fig. 12, with the black line representing the prediction error based on CUI and the blue line representing the prediction error based on MTS. It can be observed that the initial prediction error of MTS is slightly higher than that of CUI, aligning with the results of the previous single-step prediction. However, after the 15th time step, the error of CUI surpasses that of MTS and exhibits a faster growth trend. The maximum error reached throughout the entire inference process is 0.0653, approximately 1.5 times that of MTS. This discrepancy in growth trends can be attributed to the accumulation of learning errors in the neural network caused by non-physical solutions near walls. Furthermore, the modeling technique based on structured grid transformation demonstrates better capability in suppressing the divergence of flow field structure during inference compared to uniform interpolation.

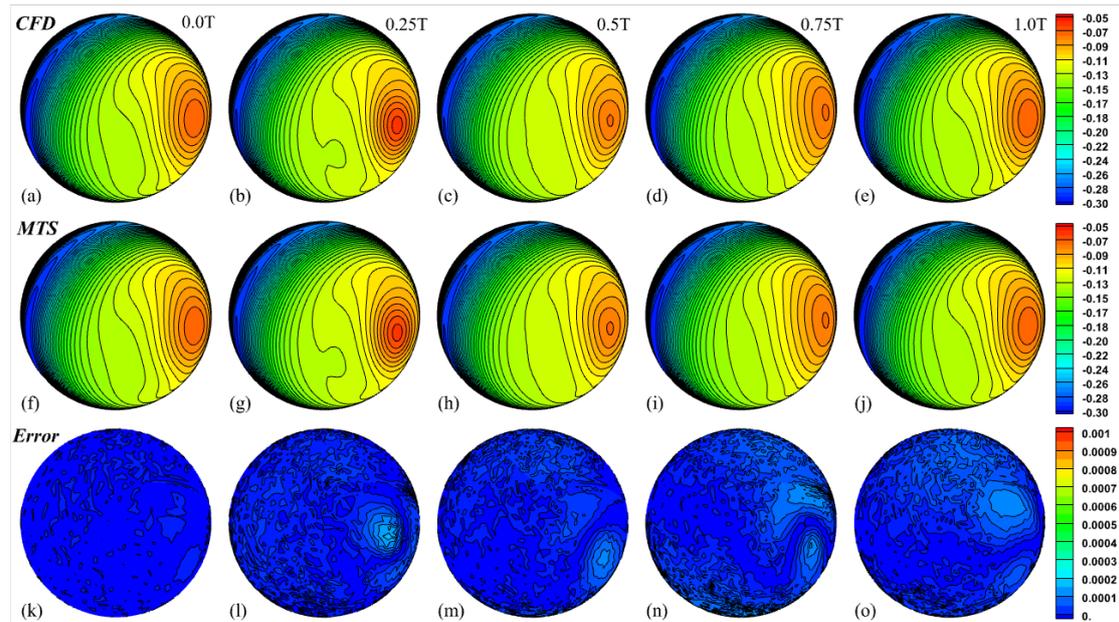

Fig. 13. Comparison of the flow field in the negative pressure region on the surface of a sphere at different time steps (0.0T, 0.25T, 0.5T, 0.75T, 1.0T) within a cycle between CFD results and MTS-based model predictions. (a)-(e) The non-dimensional pressure obtained by CFD. (f)-(j) The non-dimensional pressure obtained by the MTS-based method. (k)-(o) The absolute error between them.

In general, regardless of the dataset used, the model's final error remains within an acceptable range, providing strong evidence for the robustness of the model in the long-term prediction of unsteady flow. The detailed model prediction results based on the mesh transformation and stitching strategy are presented below.

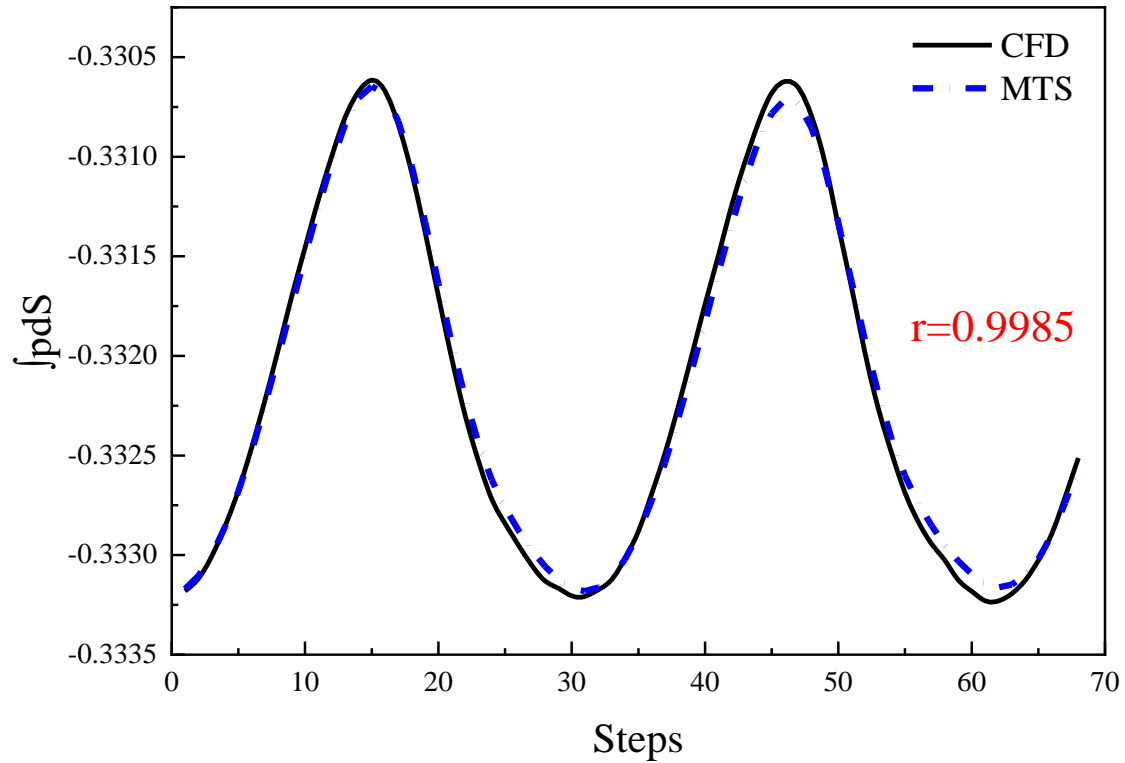

Fig. 14. The change curve of the integral of non-dimensional pressure on the spherical surface with the time step

We have successfully observed the distribution of characteristic quantities on the surface of the sphere and can determine important information, such as the position of flow separation on the sphere. In addition, understanding the dynamic evolution of this information over time is crucial for engineering applications such as structural design. Fig. 13 shows the non-dimensional pressure field obtained from both CFD and prediction, respectively, on the surface of the sphere at different time steps (0.0T, 0.25T, 0.5T, 0.75T, 1.0T) in a cycle. The fact that the flow fields at 0.0T and 1.0T are nearly identical confirms the periodic nature of the flow field. It can be seen from Fig. 13 that the pressure core region exhibits pulsating regularity, indicating the shedding of the vortex from the surface. Notably, the overall prediction error level is low, and the predictions based on MTS closely align with the CFD results. The correlation coefficient between the prediction and CFD results of the integral value of *p* on the spherical surface, as shown in Fig. 14, reaches an impressive value of 0.9985.

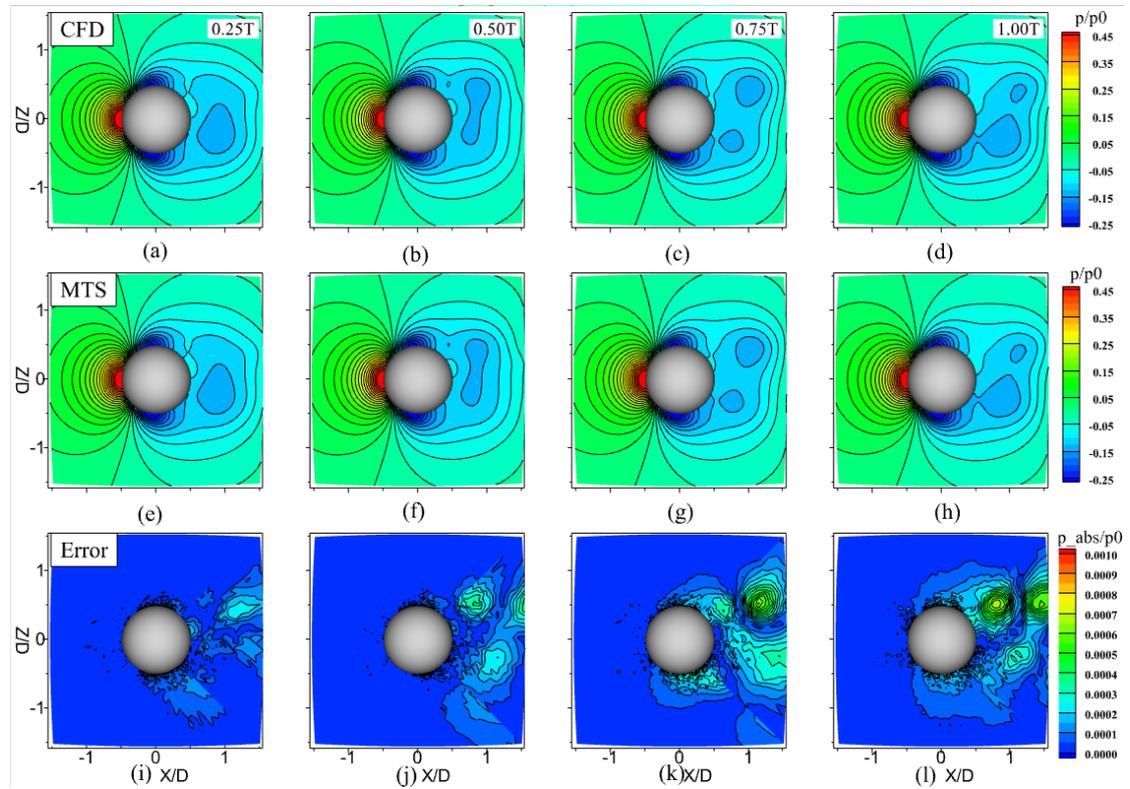

Fig. 15. Comparison of non-dimensional pressures at xz cross-section(y=0) at different time steps (0.25T, 0.5T, 0.75T, 1.0T) within a cycle: (a)-(d) CFD; (e)-(h) MTS-based method; (i)-(l) The absolute error between them.

Fig. 15 shows the non-dimensional pressure cloud diagram of the entire flow field on the xz section(y=0) at different time steps within a cycle. The neural network achieves accurate predictions of the pressure field's detailed evolution over time, with a maximum error of less than 0.001. The error is particularly lower around the sphere and is primarily concentrated in the wake region, where there are large gradient changes.

Furthermore, the model is capable of simultaneously predicting multiple features. Fig. 16 illustrates the prediction results of the velocity field and pressure field after two cycles, along with the corresponding CFD results. The error accumulation over the two cycles is minimal, and the time evolution of the physical variables in different channels exhibits similarity, which aids in the learning process of the neural network. The absolute error of the pressure field is approximately 0.001, while the maximum error of the velocity field is around 0.003. These results demonstrate the neural network's ability to faithfully predict the velocity and pressure fields in complex 3D flows.

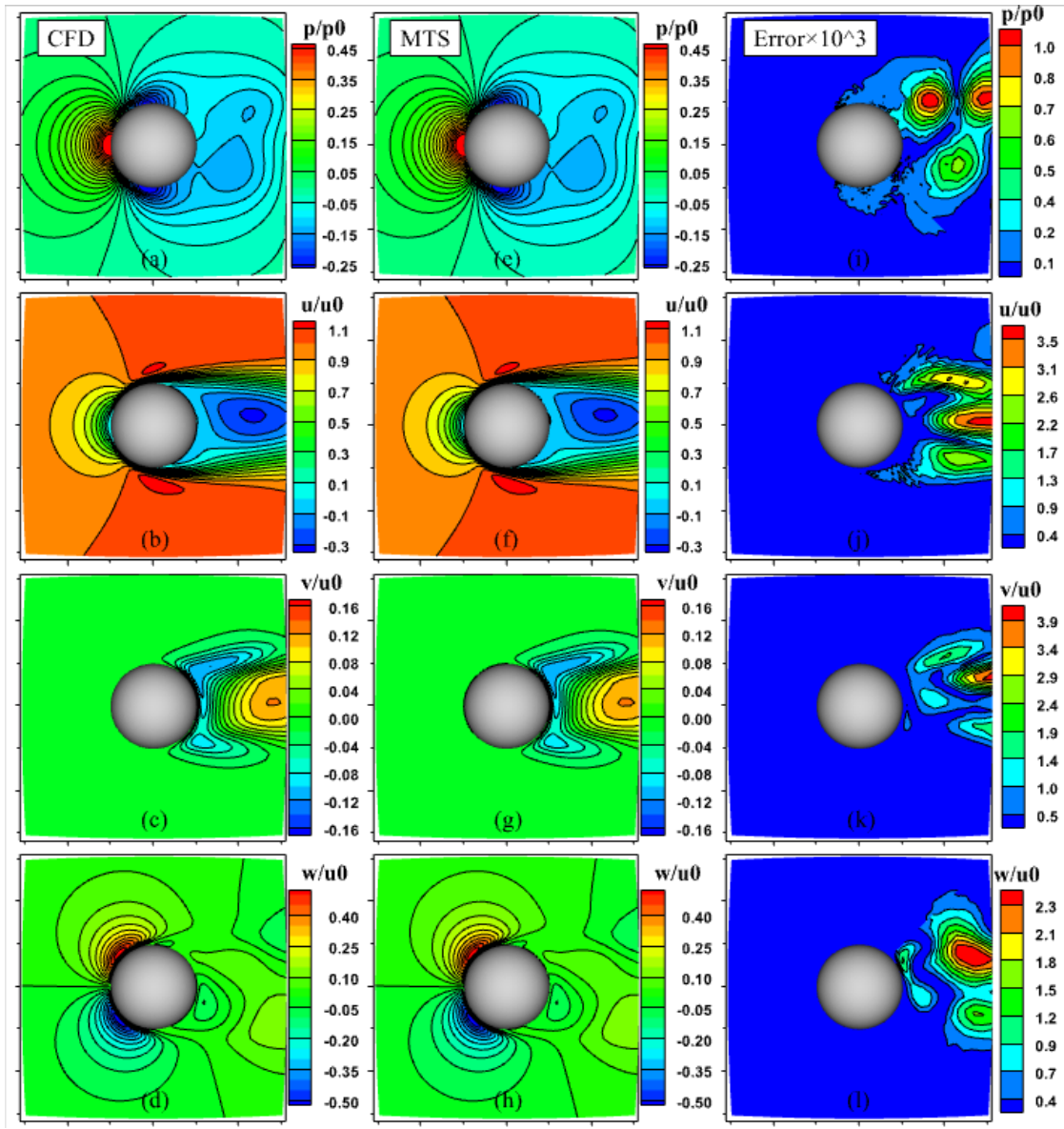

Fig. 16. Comparison of non-dimensional flow fields between model predictions and CFD results after two cycles: the left column row is the CFD result, the middle is the MTS-based forecast, and the right column is the absolute error magnified 1000 times; (a)(e)(i) pressure $p$, (b)(f)(j) streamwise velocity $u$, (c)(g)(k) normal velocity $v$, (d)(h)(l) spanwise velocity $w$.

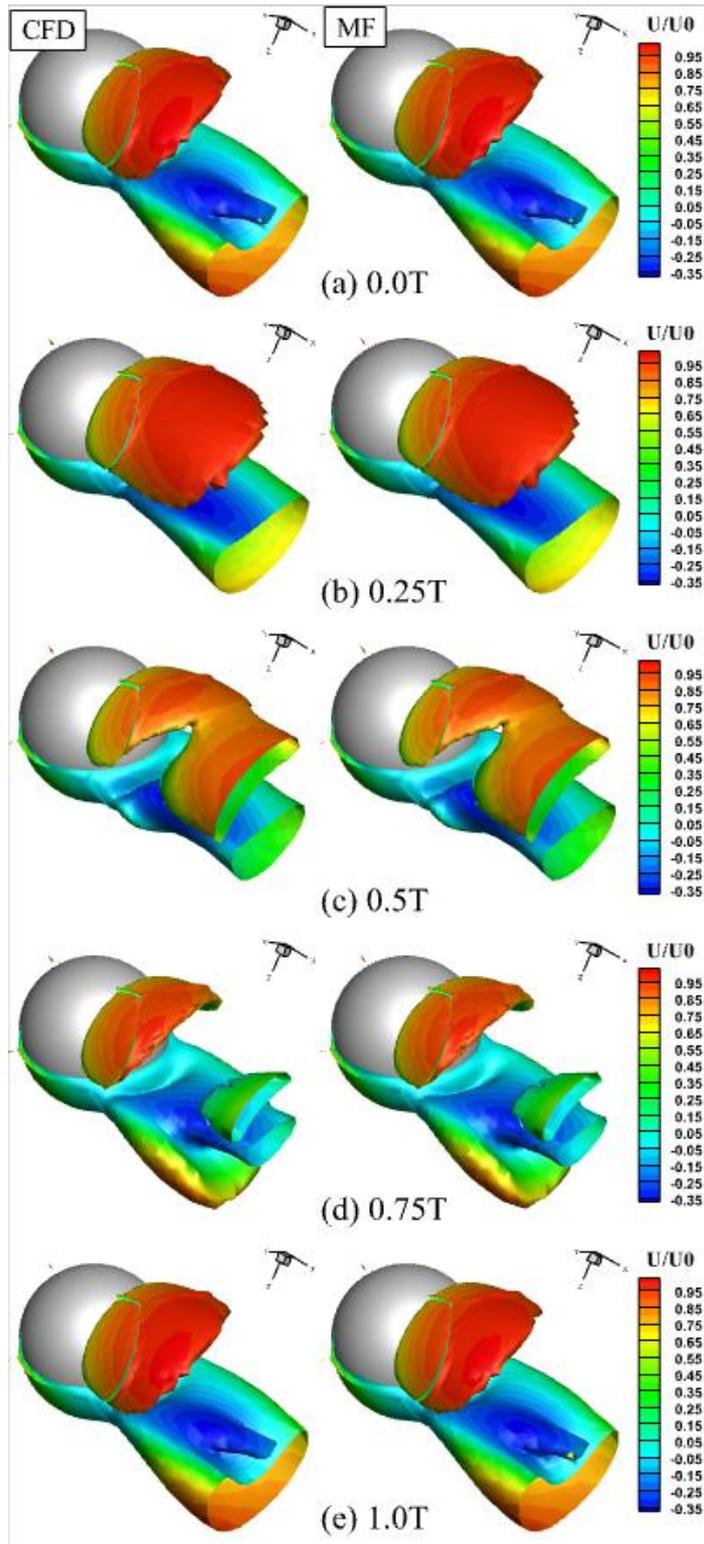

Fig. 17. Comparison of Vorticity isosurfaces at different time steps (0.25T, 0.5T, 0.75T, 1.0T) within a cycle in 3D space between model prediction and CFD. (Colors the vorticity isosurface with the value of the non-dimensional velocity field.)

To visualize the vortex structure of the three-dimensional flow field, we present the vorticity isosurface $w_x = 0.1$ at different time steps within a cycle in Fig. 17. The color on each isosurface represents the corresponding dimensionless velocity. From Fig. 17, we can observe that the direction and size of the vortex shedding align closely with the CFD results, with some acceptable differences in certain areas.

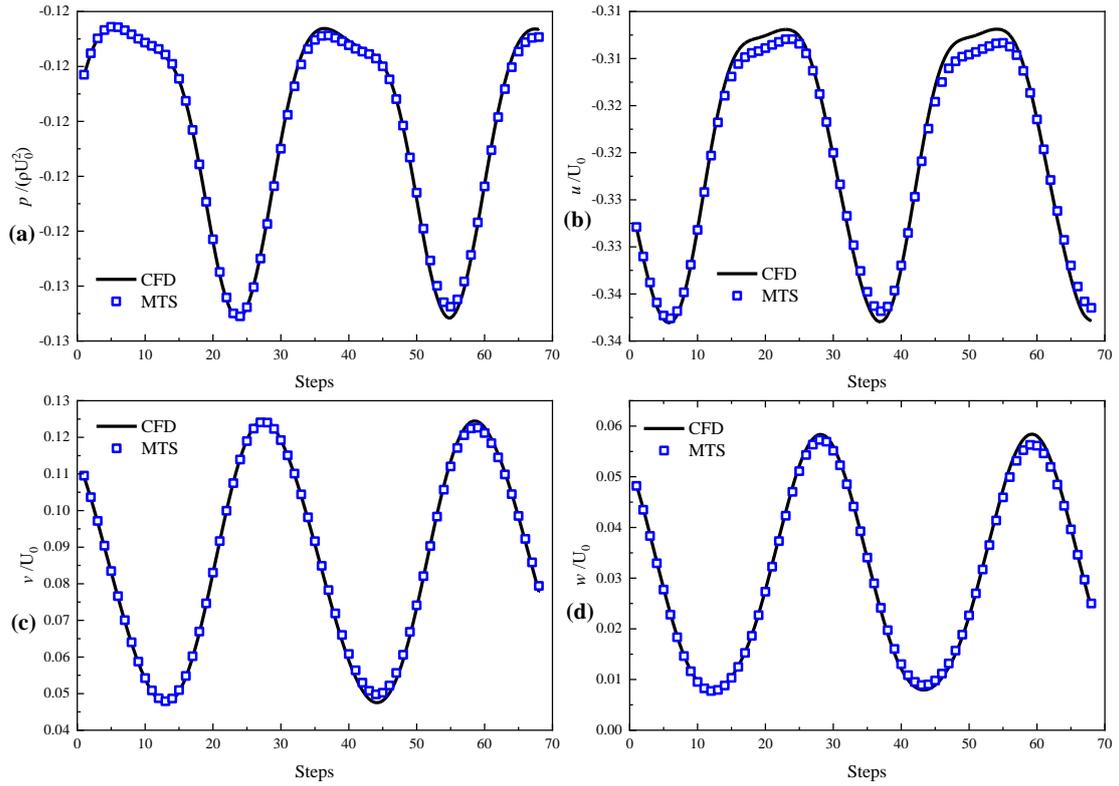

Fig. 18. The variation curves of each variable value at the feature point with the inference time step. The solid black line represents the results of CFD, and the blue square points represent the prediction results based on MTS. (a) pressure p, (b) streamwise velocity u, (c) normal velocity v, (d) spanwise velocity w.

A specific point at coordinates (1, 0.2, 0) in the flow was selected to analyze the temporal evolution of physical quantities, as depicted in Fig. 18. The plotted curves represent the non-dimensional velocity and non-dimensional pressure at that point. The results demonstrate that the MTS-based method accurately captures the periodic fluctuations of pressure and velocity. However, it slightly under-predicts the streamline velocity at the peak.

It is important to note that while the training process itself may require a significant amount of time, once trained, the ResUnet3D neural network is capable of obtaining data within one cycle in just 7 seconds. In comparison, performing a full-order CFD numerical simulation on 32 CPUs takes approximately 1 hour, resulting in an acceleration effect of over two orders of magnitude.

**D. Evaluation of the Gradient Loss Term**

In the previous chapters, the weights of the loss function were automatically adjusted using a multi-task learning adaptive strategy, resulting in good model performance. In this section, we investigate the impact of adding the gradient loss term to the neural network training and inference by setting different weight ratios for the loss functions.

To ensure reasonable weighting of the losses, we set the weight ratio $\frac{\beta}{\alpha}$ (0, 0.05, 0.1, 0.2, 1.0, 2.0, auto) between the gradient loss and the intensity loss. The ratio 0 represents pure intensity loss as a baseline, while "auto" represents an adaptive adjustment strategy. We gradually increase the weight of the gradient loss term in the baseline model, and the corresponding test error and inference error are illustrated in Fig. 19.

For single-step testing, we observe a slight increase in model accuracy with the increase in weight factor $\frac{\beta}{\alpha}$. However, when $\frac{\beta}{\alpha}$ becomes too large ($\frac{\beta}{\alpha} \geq 2$), the model's accuracy begins to deteriorate. From Fig. 19(b), it is evident that the pure intensity loss-based model is unable to effectively perform long-term inference tasks. Additionally, we have observed that the inclusion of the gradient loss term can effectively mitigate the accumulation of errors. However, it is important to note that there is a limit to increasing the weight of the gradient loss term. Moreover, the training method based on the adaptive adjustment strategy consistently maintains a high level of model accuracy, thereby reducing the necessity for extensive hyperparameter search and saving valuable time.

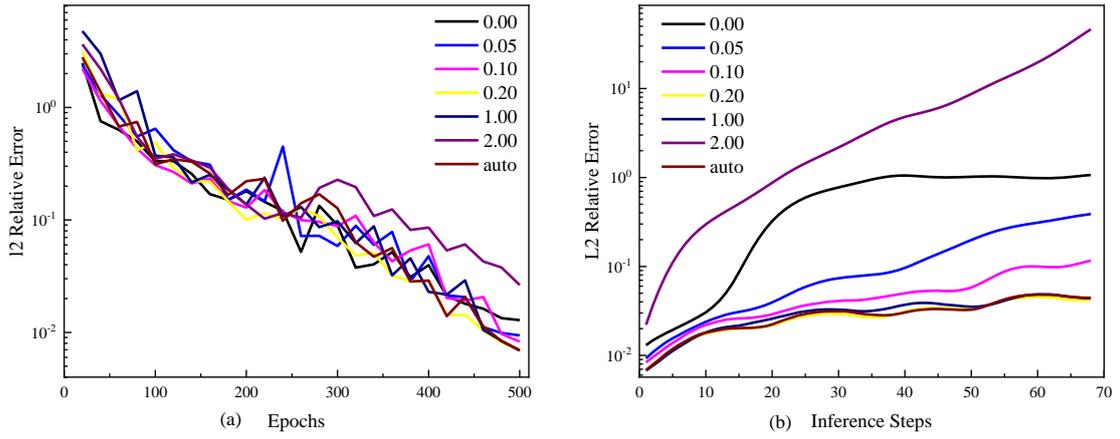

Fig. 19. L2 relative error in (a) model testing and (b) model inference under different loss weight ratios (0.00,0.05,0.10.0.20.1.00,2.00, and auto)

## V. CONCLUSIONS

A strategy based on multi-block structured grid transformation and deep learning is proposed for fast prediction of 3D unsteady flow, with a special emphasis on the accuracy of the flow near walls. Firstly, the unsteady flow is numerically simulated using the Navier-Stokes equation based on multi-block structured grids. To avoid information loss caused by interpolation, multi-block mesh transformation and stitching techniques are directly adopted to construct regular structured data. Subsequently, the ResUnet3D model based on the fully convolutional neural network is proposed to establish the nonlinear relationship between the current flow snapshot and the future flow snapshot. We verify the effectiveness of this strategy using a laminar flow around the sphere with Re=300.

To the best of our knowledge, this paper is the first attempt to predict 3D unsteady flow based on grid transformation technology. In order to demonstrate the combined advantages of MTS and deep learning, the traditional Cartesian uniform interpolation method is also used in this paper to construct regular training flow snapshots. The test results show that both methods achieve similar overall accuracy, but only the neural network based on the MTS method accurately predicts the flow details at the wall. It is worth noting that the model based on the CUI method has a faster error accumulation rate during the iterative prediction process, but the final acceptable error demonstrates the practicality of the neural network. We found that long-term stable prediction of periodic flow can be achieved based on the flow field information at the current time step. This is mainly attributed to the improved distribution of grid points and the addition of the gradient loss term to suppress the dissipation of the flow structure. In other words, the addition of the gradient loss significantly enhances the performance of the original intensity loss function, and the multi-task learning strategy can automatically determine an optimal ratio between the two. Ultimately, the MTS-based ResUnet3D model faithfully predicts physical quantities such as pressure and velocity at different time steps of the wall flow field or the overall flow field. Additionally, the neural network model significantly speeds up computations by two orders of magnitude compared to traditional CFD calculations.

However, this research is currently limited to the prediction of unsteady flow under fixed topology. While geometric information can be incorporated into neural networks, the generalization of shape is constrained to meshes with the same topology. The mesh splicing strategy, while effective in capturing near-wall features, may encounter challenges when dealing with complex shapes. And the generalization ability of the model needs to be further validated across a wider range of flow scenarios, including compressible and turbulent flows. Future research will focus on flow inference in the presence of variable shapes and operating conditions.

**ACKNOWLEDGMENTS**

This research was supported by the National Natural Science Foundation of China (Grant Nos.52192633 and 92152301)), the Natural Science Foundation of Shaanxi Province (Grant No.2022JC-03), and the National Key R&D Program of China (Grant No. 2022ZD0117804).

**AUTHOR DECLARATIONS**

**Conflict of Interest**

The authors have no conflicts to disclose.

**DATA AVAILABILITY**

The data that support the findings of this study are available from the corresponding author upon reasonable request.